# Comment on
# six papers published by M.A. El-Hakiem and his co-workers in International Communications in Heat and Mass Transfer, Journal of Magnetism and Magnetic Materials and Heat and Mass Transfer

Asterios Pantokratoras
Associate Professor of Fluid Mechanics
School of Engineering, Democritus University of Thrace,
67100 Xanthi – Greece
e-mail:apantokr@civil.duth.gr

1. "Joule heating effects on magneto hydrodynamic free convection flow of a micro polar fluid", by M.A. El-Hakiem, A.A. Mohammadein and S.M.M. El-Kabeir [**International Communications in Heat and Mass Transfer**, 26, 1999, pp. 219-227]

2. "Viscous dissipation effects on MHD free convection flow over a nonisothermal surface in a micro polar fluid", by M.A. El-Hakiem [**International Communications in Heat and Mass Transfer**, 27, 2000, pp. 581-590]

3. "Heat and mass transfer in magneto hydrodynamic flow of a micro polar fluid on a circular cylinder with uniform heat and mass flux", by M.A. Mansour, M.A. El-Hakiem and S.M. El-Kabeir [**Journal of Magnetism and Magnetic Materials**, 220, 2000, pp. 259-270]

4. "Thermal radiation effect on non-Darcy natural convection with lateral mass transfer", by M.A. El-Hakiem and M.F. El-Amin [**Heat and Mass Transfer**, 37, 2001, pp. 161-165]

5. "Mass transfer effects on the non-Newtonian fluids past a vertical plate embedded in a porous medium with non-uniform surface heat flux", by M.A. El-Hakiem and M.F. El-Amin [**Heat and Mass Transfer**, 37, 2001, pp. 293-297]

6. "Combined convection in non-Newtonian fluids along a nonisothermal vertical plate in a porous medium with lateral mass flux", by M.A. El-Hakiem [**Heat and Mass Transfer**, 37, 2001, pp. 379-385]

In all the above papers there is a problem in some figures. It is known in boundary layer theory that velocity and temperature profiles approach the ambient fluid conditions asymptotically and do not intersect the line which represents the boundary conditions. In the following figure we show schematically one correct profile and one wrong profile.

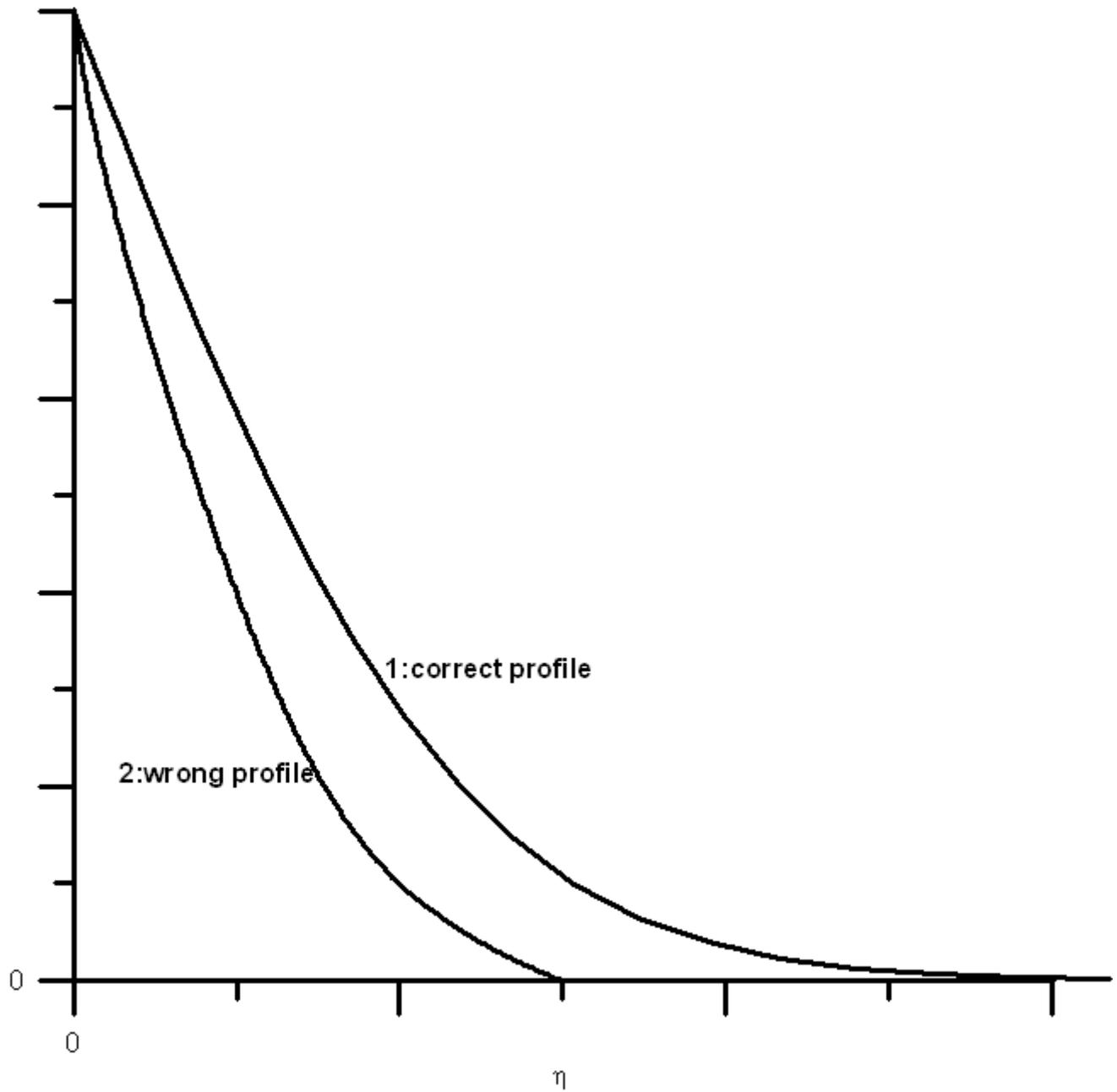

Figure1. Correct and wrong velocity or temperature profiles in boundary layer flow.

In the above six papers there are many profiles which are similar to profile 2 as follows:
First paper: All profiles included in figure 1
Second paper: All profiles included in figures 1 and 4
Third paper: All profiles included in figures 7 and 10.
Fourth paper: Some profiles included in figures 2 and 3.
Fifth paper: Some profiles included in figure 6
Sixth paper: Some profiles included in figures 3, 5, 6 and 7.

All the above profiles are probably truncated due to a small calculation domain used and are wrong.